\documentclass[pdftex,twocolumn,epjc3]{svjour3}          

\RequirePackage{amsmath}
\RequirePackage{multirow}
\RequirePackage{subfigure}
\RequirePackage{hyperref}

\allowdisplaybreaks[3]

\RequirePackage[T1]{fontenc}

\smartqed  

\RequirePackage{graphicx}

\journalname{Eur. Phys. J. C}

\begin{document}

\title{Excited $\Omega_b$ baryons and fine structure of strong interaction}

\author{Hua-Xing Chen\thanksref{addr1,addr2} \and Er-Liang Cui\thanksref{addr3} \and Atsushi Hosaka\thanksref{addr4,addr5} \and Qiang Mao\thanksref{addr6} \and Hui-Min Yang\thanksref{addr2}
}

\institute{School of Physics, Southeast University, Nanjing 210094, China\label{addr1}
\and
School of Physics, Beihang University, Beijing 100191, China\label{addr2}
\and
College of Science, Northwest A{\rm \&}F University, Yangling 712100, China\label{addr3}
\and
Research Center for Nuclear Physics (RCNP), Osaka University, Ibaraki 567-0047, Japan\label{addr4}
\and
Advanced Science Research Center, Japan Atomic Energy Agency (JAEA), Tokai 319-1195, Japan\label{addr5}
\and
Department of Electrical and Electronic Engineering, Suzhou University, Suzhou 234000, China\label{addr6}}

\date{Received: date / Accepted: date}

\maketitle

\begin{abstract}
The heavy baryon system bounded by the strong interaction has a rich internal structure, so its mass spectra can have the fine structure similar to the line spectra of atom bounded by the electromagnetic interaction. We systematically study the internal structure of $P$-wave $\Omega_b$ baryons and calculate their $D$-wave decay properties. The present study, together with our previous studies on their mass spectra and $S$-wave decay properties, suggest that all the four excited $\Omega_b$ baryons recently discovered by LHCb can be well explained as $P$-wave $\Omega_b$ baryons, and their beautiful fine structure is directly related to the rich internal structure of $P$-wave $\Omega_b$ baryons.
\end{abstract}

$\\$
{\it Introduction} ---
The electromagnetic interaction holds the electrons and protons together inside a single atom, leading to the gross, fine, and hyperfine structures of the line spectra. The strong interaction occurring between quarks and gluons is similar in some aspects, and it is interesting to investigate whether the hadron spectra also have the fine structure. An ideal platform to study this is the heavy baryon system containing one
charm or bottom quark, which is interesting in a theoretical point of view~\cite{Korner:1994nh,Manohar:2000dt,Klempt:2009pi}: the light quarks and gluons circle around the nearly static heavy quark, so that the whole system behaves as the QCD analogue of the hydrogen bounded by the electromagnetic interaction. This system has a rich internal structure, so its mass spectra can have the fine structure similar to hydrogen spectra~\cite{pdg,Copley:1979wj,Karliner:2008sv,Chen:2016spr}.

In the past years important progress has been made in this field, and many heavy baryons were observed in experiments~\cite{pdg,Aaij:2017nav,Yelton:2017qxg,Aaij:2018yqz,Aaij:2018tnn,Aaij:2020cex}. Especially, in 2017 the LHCb Collaboration discovered as many as five excited $\Omega_c$ states, $\Omega_c^0(3000)$, $\Omega_c^0(3050)$, $\Omega_c^0(3066)$, $\Omega_c^0(3090)$, and $\Omega_c^0(3119)$, simultaneously in the $\Xi_c^+ K^-$ mass spectrum~\cite{Aaij:2017nav}. Very recently, they further discovered four excited $\Omega_b$ states at the same time in the $\Xi_b^0 K^-$ mass spectrum~\cite{Aaij:2020cex}:
\begin{eqnarray}
\nonumber \Omega_b(6316)^-   &:& M = 6315.64 \pm 0.31 \pm 0.07 \pm 0.50~{\rm MeV} \, ,
\\                 && \Gamma < 2.8~{\rm MeV} \, ,
\\ \nonumber \Omega_b(6330)^-&:& M = 6330.30 \pm 0.28 \pm 0.07 \pm 0.50~{\rm MeV} \, ,
\\                 && \Gamma < 3.1 ~{\rm MeV} \, ,
\\ \nonumber \Omega_b(6340)^-&:& M = 6339.71 \pm 0.26 \pm 0.05 \pm 0.50~{\rm MeV} \, ,
\\                 && \Gamma < 1.5~{\rm MeV} \, ,
\\ \nonumber \Omega_b(6350)^-&:& M = 6349.88 \pm 0.35 \pm 0.05 \pm 0.50~{\rm MeV} \, ,
\\                 && \Gamma = 1.4^{+1.0}_{-0.8} \pm 0.1~{\rm MeV} \, .
\end{eqnarray}
These excited $\Omega_c$ and $\Omega_b$ states are good candidates of $P$-wave charmed and bottom baryons. To understand them, many phenomenological methods and models have been applied, such as
various quark models~\cite{Capstick:1986bm,Chen:2007xf,Ebert:2007nw,Garcilazo:2007eh,Zhong:2007gp,Ortega:2012cx,Yoshida:2015tia,Nagahiro:2016nsx,Kim:2017khv,Wang:2017kfr,Yang:2017rpg,Wang:2017vnc,Chen:2018vuc,Santopinto:2018ljf},
the chiral perturbation theory~\cite{Lu:2014ina,Cheng:2015naa},
the molecular model~\cite{GarciaRecio:2012db,Liang:2014eba,An:2017lwg,Montana:2017kjw,Debastiani:2017ewu,Chen:2017xat,Nieves:2017jjx,Huang:2018wgr},
Lattice QCD~\cite{Padmanath:2013bla,Padmanath:2017lng},
and QCD sum rules~\cite{Bagan:1991sg,Groote:1996em,Huang:2000tn,Wang:2003zp,Wang:2017zjw,Agaev:2017jyt,Agaev:2017ywp,Liu:2007fg,Chen:2015kpa,Mao:2015gya,Chen:2017sci,Cui:2019dzj,Yang:2019cvw}, etc.
We refer to reviews~\cite{Chen:2016spr,Cheng:2015iom,Crede:2013sze,Amhis:2019ckw} and references therein for their recent progress.

In Refs.~\cite{Chen:2015kpa,Mao:2015gya} we systematically studied mass spectra of $P$-wave heavy baryons using QCD sum rules~\cite{Shifman:1978bx,Reinders:1984sr} within the the framework of heavy quark effective theory (HQET)~\cite{Grinstein:1990mj,Eichten:1989zv,Falk:1990yz}. Later in Refs.~\cite{Chen:2017sci,Cui:2019dzj,Yang:2019cvw} we systematically studied their $S$-wave decays into ground-state heavy baryons together with light pseudoscalar and vector mesons, using light-cone sum rules~\cite{Balitsky:1989ry,Braun:1988qv,Chernyak:1990ag,Ball:1998je,Ball:2006wn} still within HQET. Recently, we have applied the same method to systematically study their $D$-wave decays into ground-state heavy baryons and light pseudoscalar mesons~\cite{dwave}. Hence, we have performed a rather complete study on both the mass spectra and strong decay properties of $P$-wave heavy baryons within the framework of HQET.

In this letter we shall apply these sum rule results to study the four excited $\Omega_b$ baryons recently observed by LHCb~\cite{Aaij:2020cex}. We shall find that all of them can be well interpreted as $P$-wave $\Omega_b$ baryons, so that both their mass spectra and decay properties can be well explained. Especially, their beautiful fine structure can be well explained in the framework of HQET, that is directly related to the rich internal structure of $P$-wave $\Omega_b$ baryons.

\begin{figure*}[]
\begin{center}
\scalebox{0.6}{\includegraphics{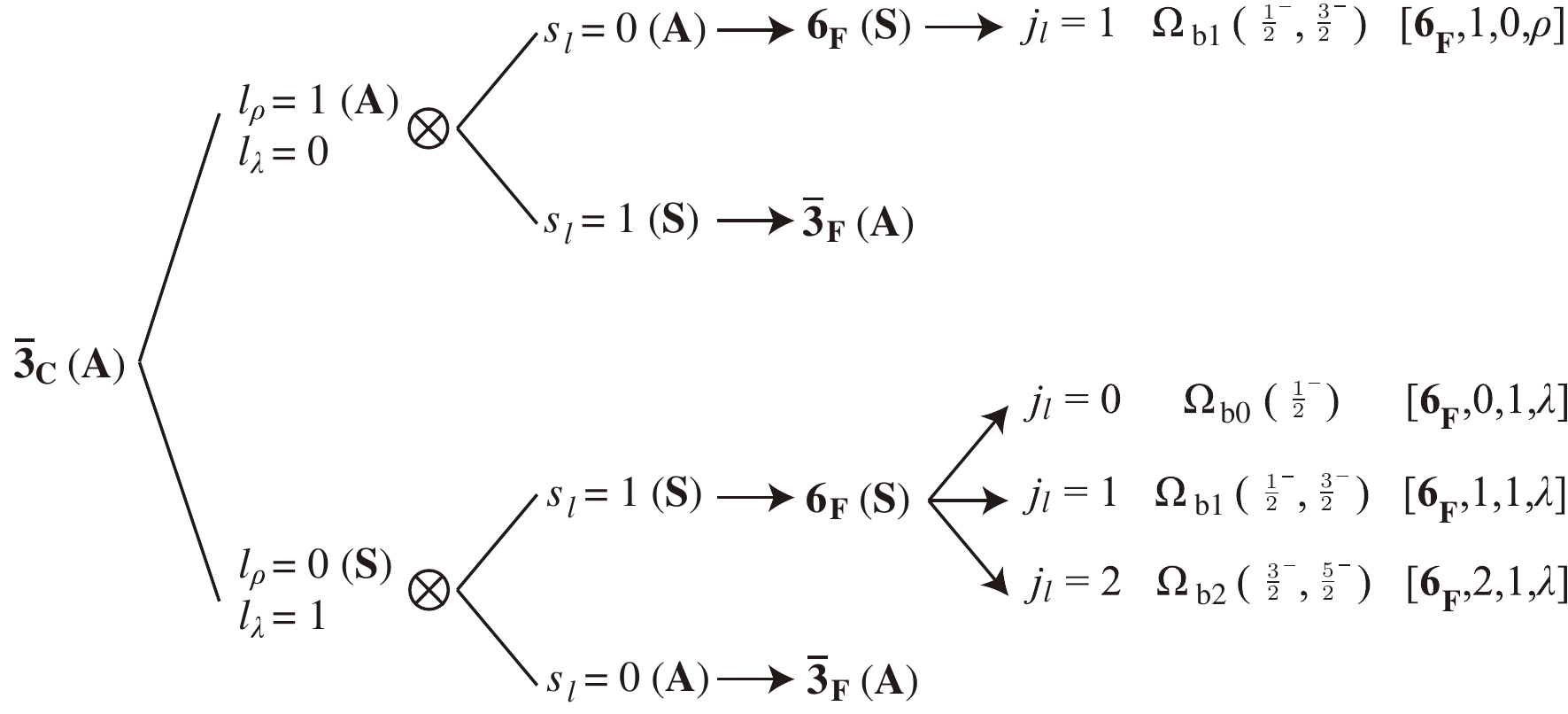}}
\caption{Categorization of $P$-wave $\Omega_b$ baryons.
\label{fig:pwave}}
\end{center}
\end{figure*}

$\\$
{\it A global picture from the heavy quark effective theory} ---
First let us briefly introduce our notations. A $P$-wave $\Omega_b$ baryon consists of one bottom quark and two strange quarks. Its orbital excitation can be either between the two strange quarks ($l_\rho = 1$) or between the bottom quark and the two-strange-quark system ($l_\lambda = 1$), so there are $\rho$-mode excited $\Omega_b$ baryons ($l_\rho = 1$ and $l_\lambda = 0$) and $\lambda$-mode ones ($l_\rho = 0$ and $l_\lambda = 1$). Altogether its internal symmetries are as follows:
\begin{itemize}

\item Color structure of the two strange quarks is antisymmetric ($\mathbf{\bar 3}_C$).

\item Flavor structure of the two strange quarks is symmetric, that is the $SU(3)$ flavor $\mathbf{6}_F$.

\item Spin structure of the two strange quarks is either antisymmetric ($s_l = 0$) or symmetric ($s_l = 1$).

\item Orbital structure of the two strange quarks is either antisymmetric ($l_\rho = 1$) or symmetric ($l_\rho = 0$).

\item Totally, the two strange quarks should be antisymmetric due to the Pauli principle.

\end{itemize}
Accordingly, we can categorize $P$-wave $\Omega_b$ baryons into four multiplets, as shown in Fig.~\ref{fig:pwave}. We denote them as $[\mathbf{6}_F, j_l, s_l, \rho/\lambda]$, where $j_l$ is the total angular momentum of the light components ($j_l = l_\lambda \otimes l_\rho \otimes s_l$). Each multiplet contains one or two $\Omega_b$ baryons, denoted as $[\Omega_b(j^P), j_l, s_l, \rho/\lambda]$, where $j^P$ are their total spin-parity quantum numbers ($j = j_l \otimes s_b = | j_l \pm 1/2 |$ with $s_b$ the bottom quark spin). Note that there are other four multiplets with the $SU(3)$ flavor $\mathbf{\bar 3}_F$, and we refer to Refs.~\cite{Chen:2007xf,Chen:2015kpa,Mao:2015gya} for more discussions.

$\\$
{\it Mass spectrum from QCD sum rules within HQET} ------
We have systematically constructed all the $P$-wave heavy baryon interpolating fields in Ref.~\cite{Chen:2015kpa}, and applied them to study the mass spectrum of $P$-wave bottom baryons in Ref.~\cite{Mao:2015gya} using the method of QCD sum rules within HQET. In this framework the $\Omega_b$ baryon belonging to the multiplet $[F, j_l, s_l, \rho/\lambda]$ has the mass:
\begin{equation}
m_{\Omega_b(j^P),j_l,s_l,\rho/\lambda} = m_b + \overline{\Lambda}_{\Omega_b,j_l,s_l,\rho/\lambda} + \delta m_{\Omega_b(j^P),j_l,s_l,\rho/\lambda} \, ,
\label{eq:mass}
\end{equation}
where $m_b$ is the bottom quark mass, $\overline{\Lambda}_{\Omega_b,j_l,s_l,\rho/\lambda} = \overline{\Lambda}_{\Omega_b(|j_l-1/2|),j_l,s_l,\rho/\lambda} = \overline{\Lambda}_{\Omega_b(j_l+1/2),j_l,s_l,\rho/\lambda}$ is the sum rule result evaluated at the leading order, and $\delta m_{\Omega_b(j^P),j_l,s_l,\rho/\lambda}$ is the sum rule result evaluated at the ${\mathcal O}(1/m_b)$ order:
\begin{eqnarray}
\label{eq:masscorrection}
&& \delta m_{\Omega_b(j^P),j_l,s_l,\rho/\lambda}
\\ \nonumber && = -\frac{1}{4m_{b}}(K_{\Omega_b,j_l,s_l,\rho/\lambda} + d_{j,j_{l}}C_{mag}\Sigma_{\Omega_b,j_l,s_l,\rho/\lambda} ) \, .
\end{eqnarray}
Here $C_{mag} = [ \alpha_s(m_b) / \alpha_s(\mu) ]^{3/\beta_0}$ with $\beta_0 = 11 - 2 n_f /3$, and the coefficient $d_{j,j_{l}}$ is
\begin{equation}
d_{j_{l}-1/2,j_{l}} = 2j_{l}+2\, ,~~~~~
d_{j_{l}+1/2,j_{l}} = -2j_{l} \, .
\end{equation}
Hence, the $\Sigma_{\Omega_b,j_l,s_l,\rho/\lambda}$ term is directly related to the mass splitting within the same multiplet. This term is usually positive, so the mass splitting within the same multiplet is also positive.

We clearly see from Eq.~(\ref{eq:mass}) that the $\Omega_b$ mass depends significantly (almost linearly) on the bottom quark mass, for which we used the $1S$ mass $m_b = 4.66 ^{+0.04}_{-0.03}$ GeV~\cite{pdg2} in Ref.~\cite{Mao:2015gya}, while the pole mass $m_b = 4.78 \pm 0.06$ GeV~\cite{pdg} and the $\overline{\rm MS}$ mass $m_b = 4.18 ^{+0.04}_{-0.03}$ GeV~\cite{pdg} are used in some other QCD sum rule studies. This suggests that there is considerable uncertainty in our results for absolute values of the masses, which prevents us from touching the nature of the four excited $\Omega_b$ baryons observed by LHCb~\cite{Aaij:2020cex}. However, the mass differences within the same doublet do not depend much on the bottom quark mass, so they are produced quite well with much less uncertainty and give more useful information.

Besides, we can extract even (much) more useful information from strong decay properties of $P$-wave $\Omega_b$ baryons. Before doing this, we slightly modify one of the free parameters in QCD sum rules, the threshold value $\omega_c$, to get a better description of the four excited $\Omega_b$ baryons' masses measured by LHCb~\cite{Aaij:2020cex}. The obtained results are summarized in Table~\ref{tab:mass}.

\begin{table*}[]
\begin{center}
\caption{Mass spectra of $P$-wave $\Omega_b$ baryons belonging to the bottom baryon multiplets $[\mathbf{6}_F, 1, 0, \rho]$, $[\mathbf{6}_F, 0, 1, \lambda]$, $[\mathbf{6}_F, 1, 1, \lambda]$, and $[\mathbf{6}_F, 2, 1, \lambda]$. There is considerable uncertainty in our results for absolute values of the masses due to their (almost linear) dependence on the bottom quark mass, but the mass differences within the same doublet do not depend much on the bottom quark mass, so they are produced quite well with much less uncertainty.}
\begin{tabular}{c | c  c  c | c  c | c | c}
\hline\hline
\multirow{2}{*}{~~~~~Multiplets~~~~~} & ~~~~$\omega_c$~~~~ & ~~~Working region~~~ & ~~~~~~~~~$\overline{\Lambda}$~~~~~~~~~ & ~~~Baryon~~~ & ~~~~~Mass~~~~~ & ~~Difference~~ & ~~~~~~~~~~$f$~~~~~~~~~~
\\                                    &        (GeV)       &         (GeV)        &                 (GeV)                  &    ($j^P$)   &      (GeV)     &       (MeV)    &      (GeV$^{4}$)
\\ \hline\hline
\multirow{2}{*}{$[\mathbf{6}_F(\Omega_b), 1, 0, \rho]$}
& \multirow{2}{*}{2.13} & \multirow{2}{*}{$0.26< T < 0.37$} & \multirow{2}{*}{$1.58^{+0.10}_{-0.08}$} & $\Omega_b(1/2^-)$ & $6.32^{+0.12}_{-0.10}$ & \multirow{2}{*}{$2.3^{+1.0}_{-0.9}$} & $0.13^{+0.03}_{-0.02}$
\\ \cline{5-6}\cline{8-8}
& & & & $\Omega_b(3/2^-)$ & $6.32^{+0.12}_{-0.10}$ & &$0.08^{+0.02}_{-0.01}$
\\ \hline
$[\mathbf{6}_F(\Omega_b), 0, 1, \lambda]$
& 2.00 & $0.27< T < 0.34$ & $1.54 \pm 0.08$ & $\Omega_b(1/2^-)$ & $6.34 \pm 0.11$ & -- & $0.13 \pm 0.03$
\\ \hline
\multirow{2}{*}{$[\mathbf{6}_F(\Omega_b), 1, 1, \lambda]$}
& \multirow{2}{*}{2.00} & \multirow{2}{*}{$0.38< T < 0.39$} & \multirow{2}{*}{$1.49 \pm 0.07$} & $\Omega_b(1/2^-)$ & $6.34^{+0.09}_{-0.08}$ & \multirow{2}{*}{$6.3^{+2.3}_{-2.1}$} & $0.12 \pm 0.02$
\\ \cline{5-6}\cline{8-8}
& & & & $\Omega_b(3/2^-)$ & $6.34^{+0.09}_{-0.08}$ & &$0.07 \pm 0.01$
\\ \hline
\multirow{2}{*}{$[\mathbf{6}_F(\Omega_b), 2, 1, \lambda]$}
& \multirow{2}{*}{2.08} & \multirow{2}{*}{$0.26< T < 0.37$} & \multirow{2}{*}{$1.53^{+0.11}_{-0.08}$} & $\Omega_b(3/2^-)$ & $6.35^{+0.13}_{-0.11}$ & \multirow{2}{*}{$10.0^{+4.6}_{-3.8}$} & $0.16^{+0.04}_{-0.03}$
\\ \cline{5-6}\cline{8-8}
& & & & $\Omega_b(5/2^-)$ & $6.36^{+0.13}_{-0.11}$ & &$0.07^{+0.02}_{-0.01}$
\\ \hline \hline
\end{tabular}
\label{tab:mass}
\end{center}
\end{table*}

\begin{table*}[]
\begin{center}
\caption{Strong decay properties of $P$-wave $\Omega_b$ baryons belonging to the bottom baryon multiplets $[\mathbf{6}_F, 1, 0, \rho]$, $[\mathbf{6}_F, 0, 1, \lambda]$, $[\mathbf{6}_F, 1, 1, \lambda]$, and $[\mathbf{6}_F, 2, 1, \lambda]$. In the third and fourth columns we show the results for the $S$- and $D$-wave decays of $P$-wave $\Omega_b$ baryons into $\Xi_c K$ (both $\Xi_b^0 K^-$ and $\Xi_b^- \bar K^0$), respectively. $A.M.F.$ means that these channels are forbidden due to the conservation of angular momentum; $K.F.$ means that these channels are kinematically forbidden; 0 means that decay widths of these channels are calculated to be zero; -- means that this channel is not calculated.}
\begin{tabular}{c | c | c  c | c || c}
\hline\hline
~~~~~Multiplets~~~~~ & ~~Baryon ($j^P$)~~ & ~~~~~~$S$-wave $\Xi_b K$~~~~~~ & ~~~~~~$D$-wave $\Xi_b K$~~~~~~ & ~~~~$\Xi_b^\prime K/\Xi_b^* K/\Xi_b K^*\cdots$~~~~ & Candidate
\\ \hline\hline
\multirow{2}{*}{$[\mathbf{6}_F(\Omega_b), 1, 0, \rho]$}
& $\Omega_b(1/2^-)$     &    0               &   0   &   $K.F.$  &  \multirow{2}{*}{$\Omega_b(6316)^-$}
\\ \cline{2-5}
& $\Omega_b(3/2^-)$     &   $A.M.F.$           &   0   &   $K.F.$
\\ \hline
$[\mathbf{6}_F(\Omega_b), 0, 1, \lambda]$
& $\Omega_b(1/2^-)$     & $\Gamma = 2800^{+3600}_{-1800}$~MeV    &  --   &   $K.F.$  &
\\ \hline
\multirow{2}{*}{$[\mathbf{6}_F(\Omega_b), 1, 1, \lambda]$}
& $\Omega_b(1/2^-)$     &    0               &   0   &   $K.F.$  &   $\Omega_b(6330)^-$
\\ \cline{2-6}
& $\Omega_b(3/2^-)$     &   $A.M.F.$           &   0   &   $K.F.$  &  $\Omega_b(6340)^-$
\\ \hline
\multirow{2}{*}{$[\mathbf{6}_F(\Omega_b), 2, 1, \lambda]$}
& $\Omega_b(3/2^-)$     &   $A.M.F.$  & $\Gamma = 4.7^{+6.1}_{-2.9}$~MeV &  $K.F.$  &  $\Omega_b(6350)^-$
\\ \cline{2-6}
& $\Omega_b(5/2^-)$     &   $A.M.F.$           &   0   &   $K.F.$  &
\\ \hline \hline
\end{tabular}
\label{tab:width}
\end{center}
\end{table*}

$\\$
{\it Decay property from light-cone sum rules within HQET}---
We have systematically studied various strong decay properties of $P$-wave heavy baryons in Refs.~\cite{Chen:2017sci,Cui:2019dzj,Yang:2019cvw} using light-cone sum rules within HQET. There are indeed a lot of decay processes that can happen. However, in the present case the only possible strong decay mode for $P$-wave $\Omega_b$ baryons is decaying into $\Xi_b K$ (given their largest mass to be the mass of the $\Omega_b(6350)^-$, so that all the other strong decay modes are kinematically forbidden). Actually, we can draw even stronger conclusions:
\begin{itemize}

\item All the $S$-wave decays of $P$-wave $\Omega_b$ baryons into ground-state heavy baryons and light pseudoscalar mesons can not happen (some of them are forbidden due to the conservation of angular momentum, while some due to the selection rules of the light components of the baryons), except
\begin{equation}
\Gamma\left([\Omega_b(1/2^-), 0, 1, \lambda] \to \Xi_b K\right) = 2800^{+3600}_{-1800}~{\rm MeV} \, .
\end{equation}
The above value is evaluated through the Lagrangian
\begin{equation}
\mathcal{L} = g~{\bar \Omega_b}(1/2^-) \Xi_b ~ K \, ,
\end{equation}
using the mass of $[\Omega_b(1/2^-), 0, 1, \lambda]$ given in Table~\ref{tab:mass}.

\item All the decays of $P$-wave $\Omega_b$ baryons into ground-state heavy baryons and light vector mesons (as intermediate states) are kinematically forbidden. 

\end{itemize}
Recently, we have systematically studied $D$-wave decays of $P$-wave heavy baryons into ground-state heavy baryons and light pseudoscalar mesons~\cite{dwave}. The results indicate:
\begin{itemize}

\item All the $D$-wave decays of $P$-wave $\Omega_b$ baryons into ground-state heavy baryons and light pseudoscalar mesons can not happen, except a) $[\Omega_b(3/2^-), 2, 1, \lambda] \to \Xi_b K$ and b) $[\Omega_b(1/2^-), 0, 1, \lambda] \to \Xi_b K$. The former channel (a) is calculated to be
\begin{equation}
\Gamma\left([\Omega_b(3/2^-), 2, 1, \lambda] \to \Xi_b K\right) = 4.7^{+6.1}_{-2.9}~{\rm MeV} \, ,
\end{equation}
using the mass of the $\Omega_b(6350)^-$ measured by LHCb~\cite{Aaij:2020cex} through the Lagrangian
\begin{equation}
\mathcal{L}^\prime = g^\prime~{\bar \Omega_b}^\mu(3/2^-) \gamma^\nu \gamma_5 \Xi_b ~ \partial_\mu\partial_\nu K \, ,
\end{equation}
The latter channel (b) is too large due to the $S$-wave nature of the decay mode.

\end{itemize}
We summarize the above decay properties in Table~\ref{tab:width}.

$\\$
{\it Excited $\Omega_b$ baryons in the heavy quark effective theory} ---
Based on Tables~\ref{tab:mass} and \ref{tab:width}, we can well understand the four excited $\Omega_b$ baryons observed by LHCb~\cite{Aaij:2020cex} as $P$-wave $\Omega_b$ baryons. There are altogether seven $P$-wave $\Omega_b$ baryons, belonging to four multiplets:
\begin{eqnarray}
\nonumber \Omega_b(1/2^-) \, , \Omega_b(3/2^-) &\in& [{\bf 6}_F, 1, 0, \rho] \, ,
\\ \nonumber \Omega_b(1/2^-) ~~~~~~~~~~~~~~~~~~\,          &\in& [{\bf 6}_F, 0, 1, \lambda] \, ,
\\ \nonumber \Omega_b(1/2^-) \, , \Omega_b(3/2^-) &\in& [{\bf 6}_F, 1, 1, \lambda] \, ,
\\ \nonumber \Omega_b(3/2^-) \, , \Omega_b(5/2^-) &\in& [{\bf 6}_F, 2, 1, \lambda] \, .
\end{eqnarray}
Our results suggest:
\begin{itemize}

\item The width of $[\Omega_b(1/2^-), 0, 1, \lambda]$ is too large for it to be observed in experiments.

\item Only the natural width of the $\Omega_b(6350)^-$ was measured by LHCb to be ``$2.5\sigma$ from zero'', that is $\Gamma_{\Omega_b(6350)^-} = 1.4^{+1.0}_{-0.8}\pm0.1$ MeV~\cite{Aaij:2020cex}. Its best candidate is $[\Omega_b(3/2^-), 2, 1, \lambda]$, whose width is calculated to be $\Gamma_{[\Omega_b(3/2^-), 2, 1, \lambda]} = 4.7^{+6.1}_{-2.9}$ MeV, quite narrow because this is a $D$-wave decay mode. The $\Omega_b(6350)^-$ is the partner state of the $\Sigma_b(6097)^\pm$~\cite{Aaij:2018tnn} and $\Xi_b(6227)^-$~\cite{Aaij:2018yqz}, and it has another partner state, $[\Omega_b(5/2^-), 2, 1, \lambda]$, whose mass is $10.0^{+4.6}_{-3.8}$~MeV larger.

\item The natural widths of the $\Omega_b(6330)^-$ and $\Omega_b(6340)^-$ were both measured by LHCb to be ``consistent with zero'', and their mass difference was measured to be about 9.4~MeV~\cite{Aaij:2020cex}. Their best candidates are $[\Omega_b(1/2^-), 1, 1, \lambda]$ and $[\Omega_b(3/2^-), 1, 1, \lambda]$ respectively, whose widths are both calculated to be zero and mass difference to be $6.3^{+2.3}_{-2.1}$~MeV. We are not sure about the reason why their decays into $\Xi_c K$ are both forbidden, but this might be related to some constrain(s) from their internal (flavor) symmetries.

\item The natural width of the $\Omega_b(6316)^-$ was also measured by LHCb to be ``consistent with zero''~\cite{Aaij:2020cex}. We can explain it as either $[\Omega_b(1/2^-), 1, 0, \rho]$ or $[\Omega_b(3/2^-), 1, 0, \rho]$. It can be further separated into two states with the mass splitting $2.3^{+1.0}_{-0.9}$~MeV. We would like to note here that this $\rho$-mode excitation is lower than the $\lambda$-mode, $[{\bf 6}_F(\Omega_b), 1, 1, \lambda]$, consistent with our previous results for their corresponding multiplets with the $SU(3)$ flavor $\mathbf{\bar 3}_F$~\cite{Chen:2015kpa,Mao:2015gya}, but in contrast to the quark model expectation~\cite{Copley:1979wj,Yoshida:2015tia}. However, this may be simply because that the mass differences between different multiplets have a considerable uncertainty in our framework, similar to absolute values of the masses, but unlike the mass differences within the same multiplet.

\item The reason is quite straightforward within the framework of HQET why the $\Omega_b(6316)^-$, $\Omega_b(6330)^-$, and $\Omega_b(6340)^-$ have natural widths ``consistent with zero'' but they can still be observed in the $\Xi_b^0 K^-$ mass spectrum~\cite{Aaij:2020cex}: the HQET is an effective theory, so the three $J=1/2^-$ $\Omega_b$ states can mix together and the three $J=3/2^-$ ones can also mix together, making it possible to observe them in the $\Xi_b K$ mass spectrum; while the HQET works quite well for the bottom system, so this mixing is not large and some of them still have very narrow widths.

\end{itemize}

$\\$
{\it Summary} ---
In the present study we have systematically studied the internal structure of $P$-wave $\Omega_b$ baryons and calculated their $D$-wave decay properties. Together with our previous studies on their mass spectra and $S$-wave decay properties~\cite{Chen:2015kpa,Mao:2015gya,Chen:2017sci,Yang:2019cvw}, we have systematically studied mass spectra and strong decay properties of $P$-wave $\Omega_b$ baryons using the methods of QCD sum rules and light-cone sum rules within the framework of heavy quark effective theory. Although there is considerable uncertainty in our results for absolute values of the masses due to their (almost linear) dependence on the bottom quark mass, the mass differences within the same doublet as well as strong decay properties of $P$-wave $\Omega_b$ baryons are both useful information, based on which we can well understand the four excited $\Omega_b$ baryons recently discovered by LHCb~\cite{Aaij:2020cex} as $P$-wave $\Omega_b$ baryons.

Our results suggest: the $\Omega_b(6350)^-$ is a $P$-wave $\Omega_b$ baryon with $J^P = 3/2^-$ and $\lambda$-mode excitation, and it has a $J^P = 5/2^-$ partner whose mass is $10.0^{+4.6}_{-3.8}$~MeV larger; the $\Omega_b(6330)^-$ and $\Omega_b(6340)^-$ are partner states both with $\lambda$-mode excitation, and they have $J^P = 1/2^-$ and $3/2^-$, respectively; the $\Omega_b(6316)^-$ is a $P$-wave $\Omega_b$ baryon of either $J^P = 1/2^-$ or $3/2^-$, with $\rho$-mode excitation, and it can be further separated into two states with the mass splitting $2.3^{+1.0}_{-0.9}$~MeV. The internal quantum numbers (and so internal structures) of these four excited $\Omega_b$ baryons have also been extracted, as discussed above.

The above conclusions are drawn by combining our systematical studies of mass spectra (as well as mass splittings with the same multiplets) and decay properties of $P$-wave $\Omega_b$ baryons. We would like to note that these are just possible explanations, and there exist many other possibilities for the four excited $\Omega_b$ baryons observed by LHCb~\cite{Aaij:2020cex}, so further experimental and theoretical studies are demanded to fully understand them. However, their beautiful fine structure is in any case directly related to the rich internal structure of $P$-wave $\Omega_b$ baryons. Recalling that the development of quantum physics is sometimes closely related to the better understanding of the gross, fine, and hyperfine structures of atom (hydrogen) spectra, one naturally guesses that the currently undergoing studies on heavy baryons would not only improve our understandings on their internal structures, but also enrich our knowledge of the quantum physics.

\begin{acknowledgements}
This project is supported by
the National Natural Science Foundation of China under Grant No.~11722540,
the Fundamental Research Funds for the Central Universities,
Grants-in-Aid for Scientific Research (No.~JP17K05441 (C)),
Grants-in-Aid for Scientific Research on Innovative Areas (No.~18H05407),
and
the Foundation for Young Talents in College of Anhui Province (Grant No.~gxyq2018103).
\end{acknowledgements}

%

%

\end{document}